\documentclass[journal,twoside,web]{ieeecolor}
\usepackage{tmi}
\usepackage{cite}
\usepackage{amsmath,amssymb,amsfonts}
\usepackage{algorithmic}
\usepackage{graphicx}
\usepackage{textcomp}
\usepackage{fixltx2e}
\usepackage{hyperref}

\def\BibTeX{{\rm B\kern-.05em{\sc i\kern-.025em b}\kern-.08em
    T\kern-.1667em\lower.7ex\hbox{E}\kern-.125emX}}
\usepackage[capitalize]{cleveref}
\crefname{section}{Sec.}{Secs.}
\Crefname{section}{Section}{Sections}
\Crefname{table}{Table}{Tables}
\crefname{table}{Tab.}{Tabs.}

\usepackage{fdl}

\begin{document}
\title{A Laplacian Pyramid Based Generative H\&E Stain Augmentation Network}
\author{Fangda Li, Zhiqiang Hu, Wen Chen and Avinash Kak
\thanks{This work has been submitted to the IEEE for possible publication. Copyright may be transferred without notice, after which this version may no longer be accessible.}
\thanks{Fangda Li and Avinash Kak are with Electrical and Computer Engineering at Purdue University, West Lafayatte, 47906, US (e-mail: li1208@purdue.edu; kak@purdue.edu).}
\thanks{Zhiqiang Hu and Wen Chen are with SenseTime Research, Beijing, 100080, China (e-mail: huzhiqiang@sensetime.com; chenwen@sensetime.com).}
}
\maketitle

\begin{abstract}
Hematoxylin and Eosin (H\&E) staining is a widely used sample preparation procedure for enhancing the saturation of tissue sections and the contrast between nuclei and cytoplasm in histology images for medical diagnostics.  
However, various factors, such as the differences in the reagents used, result in high variability in the colors of the stains actually recorded.  
This variability poses a challenge in achieving generalization for machine-learning based computer-aided diagnostic tools.  
To desensitize the learned models to stain variations, we propose the Generative Stain Augmentation Network (G-SAN) -- a GAN-based framework that augments a collection of cell images with simulated yet realistic stain variations.
At its core, G-SAN uses a novel and highly computationally efficient Laplacian Pyramid (LP) based generator architecture, that is capable of disentangling stain from cell morphology.
Through the task of patch classification and nucleus segmentation, we show that using G-SAN-augmented training data provides on average 15.7\% improvement in F1 score and 7.3\% improvement in panoptic quality, respectively. 
Our code is available at \url{https://github.com/lifangda01/GSAN-Demo}.
\end{abstract}


\section{Introduction}
\label{sec:intro}

Histology refers to the study of tissues and their structures through microscopic anatomy and is widely used in medical diagnosis, especially oncology.  
Due to the fact that most cells are colorless and transparent in a bright field, tissue samples must go through a routine staining process before observation under a microscope. 
The gold standard for staining uses a combination of two dyes -- Hematoxylin and Eosin (H\&E) -- mainly owing to their relatively high color consistency and ease of application.  
The former, hematoxylin, binds strongly to the DNA and RNA in the nuclei and paints them purplish blue, whereas the latter, eosin, binds to the proteins commonly found in the cytoplasmic and extracellular regions and paints them pink.

\begin{figure}[t]
\centering
\includegraphics[width=0.4\textwidth]{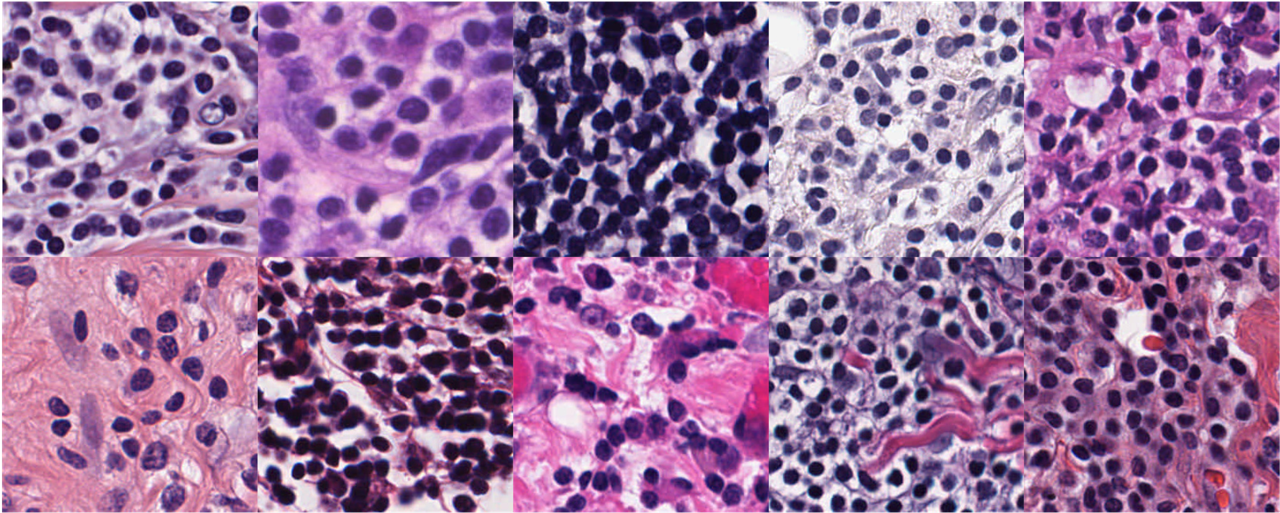}
\caption{
The high variability of H\&E-staining effects. 
The patches were extracted from different breast tissue sections that were separately stained.
}
\label{fig:variability}
\end{figure}

Despite its wide adoption, the detailed process of H\&E staining is not standardized across laboratories.  Depending on a host of factors, such as the differences in the reagents used, specific operating procedures and properties of the imaging instruments, etc., the final appearance of H\&E staining can vary significantly from slide to slide.  The patches shown in \cref{fig:variability} visually demonstrate typical examples of this phenomenon.  While this high variability in the H\&E-staining effects has been a well-known challenge for pathologists, it has also emerged as an issue in the context of computational pathology.

One of the biggest challenges for the machine learning algorithms for computational pathology is the paucity of the groundtruthed training data -- a paucity that is exacerbated by the variability in the stains.  Consider, for example, the data requirements of the algorithms for nucleus segmentation. The training data for such algorithms is scarce for two reasons: (1) it requires some domain expertise to discern the boundaries of the nuclei and the cytoplasm regions; and (2) the tediousness of manual annotation of the cell images.  And, given the data that is currently available, what reduces its effectiveness is the variability in the stains which results in overfitting and poor generalization of the machine-learning models, especially if there exist potentially unseen stains at test time.

Obviously, in order to make the most of the data that is available, what we need are strategies for desensitizing the learned models to the variability in the stains. 
Previous attempts at such model desensitization have consisted of what has come to be known as \textit{stain normalization}.  
Stain normalization alters the stain color on a pixel-by-pixel basis so that the color profile in the normalized image corresponds to a pre-specified template. 
Such normalization is applied during both training and testing. 
That is, models are trained and tested only on stain-normalized images. 
Earlier methods for stain normalization are stain-matrix based \cite{ruifrok2001quantification,macenko2009method,vahadane2016structure,xu2015sparse} and the more recent approaches leverage Convolutional Neural Networks (CNNs) \cite{tellez2019quantifying,bug2017context,shaban2019staingan,zanjani2018stain,de2021residual,shrivastava2021self,mahapatra2020structure,liang2020stain,cong2021semi}.

While stain normalization as described above is effective in reducing the stain variability, it has three significant drawbacks: (1) The extra image preprocessing steps needed at test time for stain normalization result in additional computational overhead, especially given the very large image sizes involved in histological studies.  
(2) The normalization process may involve the computationally expensive step of Sparse Non-negative Matrix Factorization (SNMF) \cite{xu2015sparse,vahadane2016structure}. 
And (3) From the standpoint of what is needed for creating models with greater generalization power, a model trained on stain-normalized images is likely to lack intrinsic versatility against stain variations, which puts the model at a higher risk of overfitting to the data.  
As a result, more recently, researchers have begun pursuing \textit{stain augmentation} in place of stain normalization for the same benefits but with the expectation of creating more generalizable models.

\begin{figure}[t]
\centering
\includegraphics[width=0.4\textwidth]{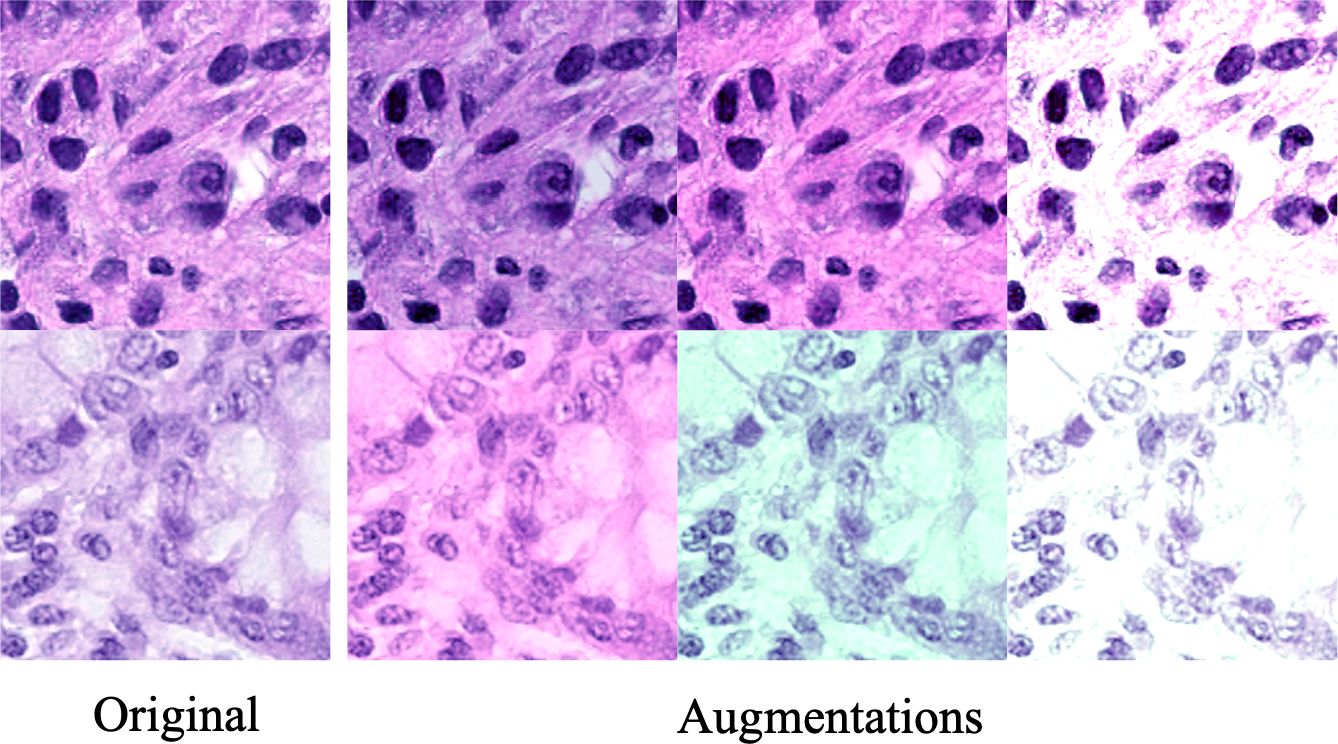}
\caption{
Jittering based augmentations created from the two original images in the left column.  
As depicted in the second row, this approach is prone to generating unrealistic stain appearances.}
\label{fig:jittering}
\end{figure}

With stain augmentation, one seeks to augment the training data with all practically possible stain variations so that a learned model possesses maximal generalizability with regard to stains. 
The effectiveness of using stain-augmented training images has been demonstrated for patch-based classification where, on the average, it led to a 40\% improvement in AUC \cite{tellez2019quantifying}. 
These authors used channel-wise color perturbation for stain augmentation.
Its idea is straightforward: One first maps the input image to an alternative color space (\eg HSV or HED using a predefined stain-matrix), then injects both multiplicative and additive random noise independently into each of the channels before reprojecting them back to RGB. 
This simple jittering-based operation is computationally efficient and was shown to be effective by the experimental results in \cite{faryna2021tailoring,wagner2021structure,scalbert2022test,shen2022randstainna}.
However, one major drawback of such a simple approach is that it is prone to generating unrealistically stained images, as illustrated in \cref{fig:jittering}.
Consequently, using HED-jittering as the only stain augmentation might not fully address the domain gap between the training and testing data, according to \cite{scalbert2022test}.

On account of the above-mentioned shortcoming of the channel-wise color perturbation approach, the focus of the ongoing research in stain augmentation has shifted to using GAN-based image-to-image translation frameworks. 
Such a framework can be used to provide either training-time stain augmentations as in the DRIT++ based HistAuGAN \cite{wagner2021structure}, the StarGAN-based framework in \cite{vasiljevic2021towards}, and the StarGANV2-based framework in \cite{scalbert2022image}, or test-time augmentations (TTAs) as in the StarGANV2-based framework in \cite{scalbert2022test}.
With its impressive data modeling capabilities, a GAN-based framework can effectively learn the distribution of the realistic stains in a high-dimensional space and subsequently create new instances of cell images with synthesized yet realistic stains obtained by sampling the learned distribution.

Despite their success, there are two main drawbacks to the existing GAN-based stain transfer or stain augmentation approaches. First, the aforementioned frameworks all group training images by their laboratory IDs and use the IDs as domain labels for training \cite{wagner2021structure,scalbert2022test,vasiljevic2021towards,scalbert2022image}. 
While such information is necessary for training multi-domain GAN frameworks, dependency on domain labels can result in frameworks that are less generalized. 
This is reflected by the fact that requiring domain-related information (\eg laboratory and organ of origin) limits the availability of training data.  
In contrast, we assume that all possible H\&E stain appearances are from a single domain.  
Together they form a single distribution and that the distribution can be sufficiently modeled by a unit Gaussian in a high-dimensional latent space. 
This independence of domain information helps G-SAN achieve better generalizability since without any domain information needed, a more diverse set of images, in terms of both tissue morphology and stain, can be used in training.

The second drawback is in regard to the computational efficiency. When used during the training or testing of a downstream task-specific model, it is important for any image augmentation algorithm to be computationally efficient.  This is especially the case in histology applications where tissue slides can have very large sizes.  Existing approaches that are based on general-purpose GAN architectures for performing stain transfer are not optimized in terms of speed.

To address the two aforementioned limitations, we propose a GAN-based stain augmentation framework that utilizes a novel generator architecture for stain transfer and the concepts of disentanglement learning. 
Our proposed generator architecture is based on the Laplacian Pyramid (LP) representation of images for ensuring that the stain transfers are structure preserving.  
More specifically, G-SAN uses the computationally heavier convolutional modules only on the low-resolution residual images of the LP, where the differences between stains are the most significant.  
As for the higher-resolution band-pass images of the LP, which capture mostly high-frequency spatial details rather than stain appearances, they are only fine-tuned by light-weight convolutional modules to both retain the structural details and to improve computational efficiency.

The G-SAN framework uses the principles of content-style disentanglement to learn to extract two independent representations from an input image: the cell morphology as content and the stain profile as style. 
Subsequently, by combining stain representations either extracted from other images or sampled stochastically, with the morphology representation from an input cell image, G-SAN can virtually re-stain the input image without altering the underlying cell structures.

We trained G-SAN in an entirely unsupervised manner, in contrast to previous works that used domain labels.  
As we demonstrate in this paper, using H\&E-stained histology images collected from a diverse set of sources for training gives G-SAN the generalization abilities with regard to both the stain appearance and the cell morphology.
The quantitative validation of our approach consists of demonstrating the effectiveness of the stain augmentations produced by G-SAN through two common downstream tasks: tumor classification and nuclei segmentation. 
For the former, the stain augmentations must help the model overcome the large domain gaps that exist between the training and testing data. 
And for the latter, the stain augmentations must be structure-preserving since any undesired modification to the underlying cell morphology would be highly punishing. 
By using our stain augmentation method, we show that the trained task-specific networks are more robust towards stain variations compared to using the current state-of-the-art in stain augmentation.

\section{Related Literature}
\label{sec:related}

\subsection{GAN-Based Stain Transfer}
\label{subsec:gan_stain_transfer}
Recent advances in GANs (Generative Adversarial Networks) have inspired several GAN-based approaches to H\&E stain-related image-to-image translation.
Using conditional generators, there now exist frameworks \cite{bug2017context,zanjani2018stain,liang2020stain,cong2021semi,tellez2019quantifying,fan2022fast} that can transform images from one or multiple stain domains into a given target stain domain.
Additionally, the success of CycleGAN \cite{zhu2017unpaired} in achieving unsupervised domain transfer has led to the development of frameworks that use cycle consistency for achieving one-to-one and many-to-one stain normalization \cite{shaban2019staingan,de2021residual,shrivastava2021self,mahapatra2020structure,vasiljevic2021towards}.
Going beyond stain normalization, frameworks that are capable of performing stain transfer among multiple stain domains have also been proposed.
Examples include the DRIT++ based 
HistAuGAN \cite{wagner2021structure}, the StarGAN-based 
framework in \cite{vasiljevic2021towards} and the StarGANV2-based 
frameworks in \cite{scalbert2022test,scalbert2022image}.
Our work is most similar to these frameworks on multi-domain stain transfer.
However, instead of defining multiple distinct stain domains commonly based on their laboratory of origin, we treat the complete set of realistic stain appearances as if coming from a single domain.  

\subsection{CNNs with Laplacian Pyramid}
\label{subsec:cnn_with_lp}
One of our important contributions in this work is the use of the Laplacian Pyramid for a highly computationally efficient yet structure-preserving CNN architecture designed specifically for H\&E stain transfer.
The method of Laplacian Pyramid decomposes an input image into a set of band-pass images, spaced an octave apart, plus a low-frequency residual.  
The popularity of this approach can be gauged by the fact that it has recently been incorporated in deep learning frameworks for various applications such as image generation \cite{denton2015deep}, image style transfer \cite{lin2021drafting}, image super-resolution \cite{lai2017deep}, etc.  
The hierarchical nature of the LP representation lends itself well to creating solutions that require adaptation to image details at different levels in the scale space.
Our LP-based generator architecture is partially inspired by the LPTN framework proposed in \cite{liang2021high}. 
More specifically, we have adopted from that work the idea of fine-tuning only the structure-rich high-resolution band-pass images with light-weight modules. 
This helps our framework preserve the spatial details in the images and, at the same time, achieve highly competitive computational efficiency.

\subsection{Learning Disentangled Representations}
We approach the modeling of the stain variability through learning to extract the following disentangled representations from an input histological image: a morphology-independent stain vector and the underlying structural representation. Our framework's learning to extract such disentangled representations is inspired by the multi-domain image-to-image translation frameworks such as those reported in \cite{lee2020drit++,choi2018stargan}.  Generally, these frameworks assume that an image can be decomposed into a domain-invariant representation and a domain-dependent representation. By enforcing the constraint that the former representation can be shared across domains, certain properties can be kept consistent through both inter- and intra-domain mappings, such as the structure of the objects.  Along similar lines, we disentangle the cell morphology, which is the stain-invariant representation in our case, from the stain representation, so that the cell structure in the images is kept consistent during stain transfer.

We summarize the stain information in the affine parameters of the learned features in the normalization layers of the generator.  Consequently, by manipulating the normalization parameters through Adaptive Instance Normalization (AdaIN) \cite{huang2017arbitrary}, we can effectively modify the stain appearance in the synthesis.  
We train this normalization-based style transfer architecture with several disentanglement-promoting learning criteria, such as the cycle-consistency loss \cite{zhu2017unpaired}, which encourages the reversibility of the learned disentangled representations, and the latent reconstruction loss \cite{zhu2017toward} that ensures the reproducibility of the disentanglement.  Subsequently, by combining arbitrary stains with the morphology representation from a given input cell image, G-SAN can generate an augmented version of the image with a simulated yet realistic looking stain.  To the best of our knowledge, G-SAN is the first CNN framework that achieves stain transfer between arbitrary H\&E stains.

\section{The Proposed G-SAN Framework}
In this section, we start with an overview of the concept of Laplacian Pyramid (LP).  
This is followed by a detailed explanation of our multi-pathway G-SAN generator architecture, which is optimized for high-resolution structure-preserving stain transfer. 
We describe the necessary design elements in our model that lead to the disentanglement of morphology and stain.  
Then, we demonstrate how the G-SAN architecture can leverage the multi-scale nature of LP in both training and inference.  
Lastly, we present the complete training procedure of our framework along with the losses used.

\subsection{The Laplacian Pyramid}
\label{subsec:lp}

\begin{figure}[t]
\centering
\includegraphics[width=0.49\textwidth]{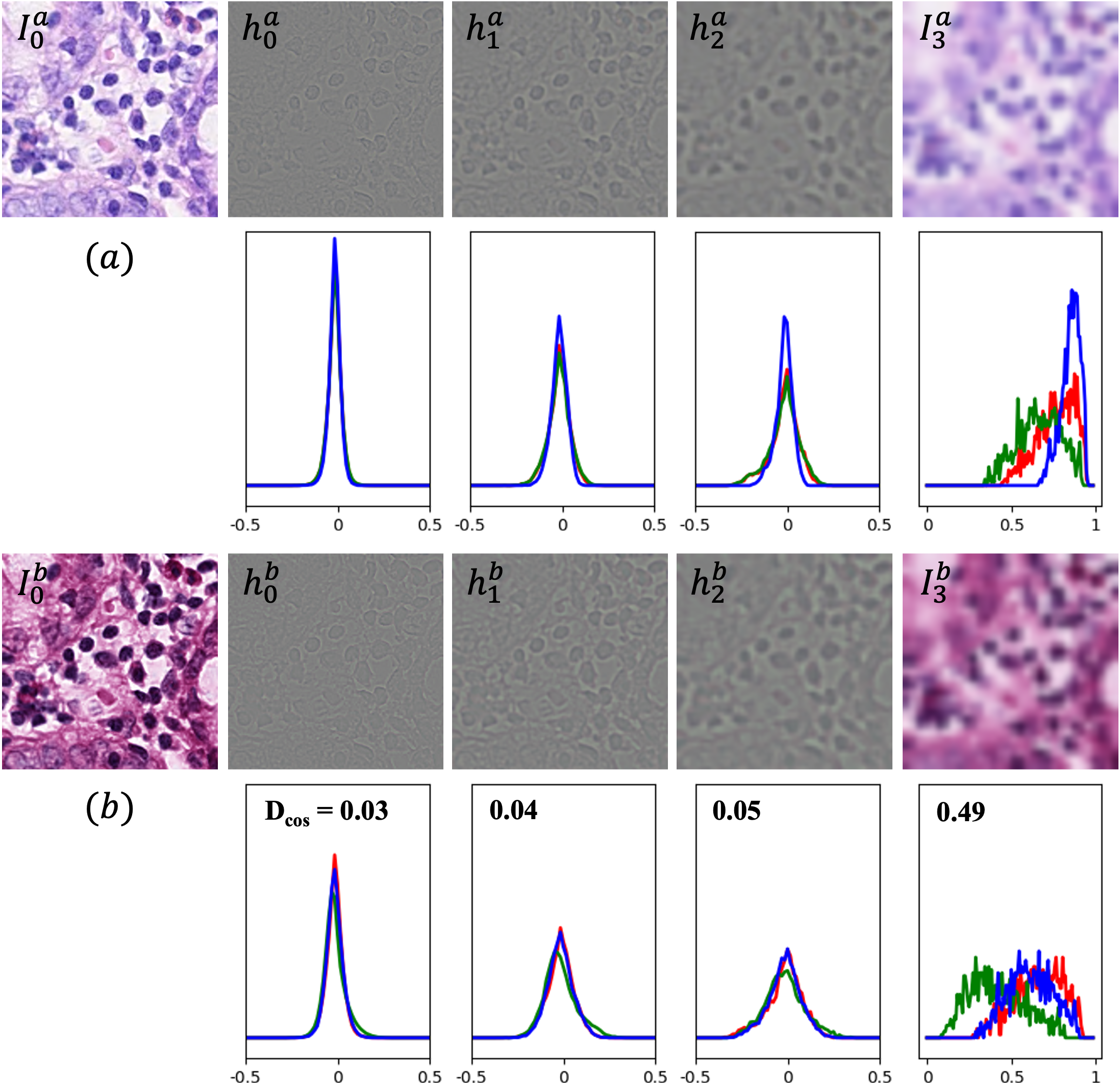}
\caption{
The Laplacian Pyramid representations with $K = 3$ of the same cell morphology with two different stains, in (a) and (b), and their RGB histograms. 
$D_{\cos}$ measures the cosine distance between the histograms of corresponding LP representations of the two images. 
While the color difference is the most prominent between the low-resolution residual images $\mI_3$, it is also evident among the high-frequency band-pass images $\vh_{k = 0,1,2}$ albeit decreasingly as the resolution increases from right to left.
Note that in the figure, the $\mI_3$ and $\vh_{k = 1,2}$ images have been up-sized to fit the display grid.
Please zoom in to get a better sense of the structures retained in the band-pass images $h_{k=0,1,2}$.
}
\label{fig:histogram}
\end{figure}

The Laplacian Pyramid is a multi-scale image representation that consists of a set of band-pass images, spaced an octave apart, and a low-resolution residual image. 
The set of band-pass images contains spatial details at consecutive frequency intervals, while the residual image is a Gaussian-blurred and downsampled version of the original input image.

To formally define the Laplacian Pyramid (LP), let $K$ denote the max image level in the LP, $g(\cdot)$ the function that convolves an image with a Gaussian kernel, and $f_{\downarrow2}(\cdot)$ / $f_{\uparrow2}(\cdot)$ the image downsampling / upsampling by 2 function, respectively.  
Then the Gaussian Pyramid (GP) of an input image $\mI$ can be written as $G(\mI) = [\mI_0, \mI_1, ..., \mI_K]$, where $\mI_0$ is the input image itself and $\mI_{k+1} = f_{\downarrow2}(g(\mI_k))$.
On the other hand, the LP of an image comprises two parts: a set of band-pass images at level 0 to $K-1$, and a residual image at level $K$.
To explain, with the definition of GP, we can first write the band-pass image of the LP at level $k = 0, ..., K-1$ as the difference between the GP image at level $k$ and the upsampled version of the GP image at level $k+1$:
\begin{equation}
\vh_k = \mI_k - f_{\uparrow2}(\mI_{k+1}).
\end{equation}
Subsequently, at the $K$th level of the LP is the low-resolution residual image, taken directly from the GP at level $K$: $L_K (\mI) = \mI_K$. 
Finally, we can now denote the complete LP representation as $L(\mI) = [\vh_0, ..., \vh_{K-1}, \mI_K]$ (examples shown in \cref{fig:histogram}). 
It is important to note that the LP decomposition of an image is lossless and fully reversible using the following backward recurrence:
\begin{equation}
\label{eq:lp-recon}
\mI_k = \vh_k + f_{\uparrow2}(\mI_{k+1}),
\end{equation}
where $\mI_0$ is the original input image.

The hierarchical separation of the high-frequency spatial details from the low-frequency residual image by the LP lends itself well to the task of stain transfer. 
Based on the observation that the stain difference between any two given input images is most prominent between the residual images $\mI_K$, as shown in \cref{fig:histogram}, G-SAN adopts an adaptive strategy that depends on the level in the LP pyramid. 
More specifically, in G-SAN, heavy convolutional modules are only allocated for translating the low-resolution residual images. 
While for the higher-resolution band-pass images, G-SAN uses light-weight convolutional modules only to fine-tune the images.
In this manner, G-SAN preserves rich spatial details in the images.
As a result, the computational burden related to the processing of the higher-resolution constituents of the images is greatly reduced while conforming to the structure-preserving needs required for stain transfer.

\subsection{The G-SAN Architecture}
\label{subsec:architecture}

\begin{figure*}[t]
\centering
\includegraphics[width=0.9\textwidth]{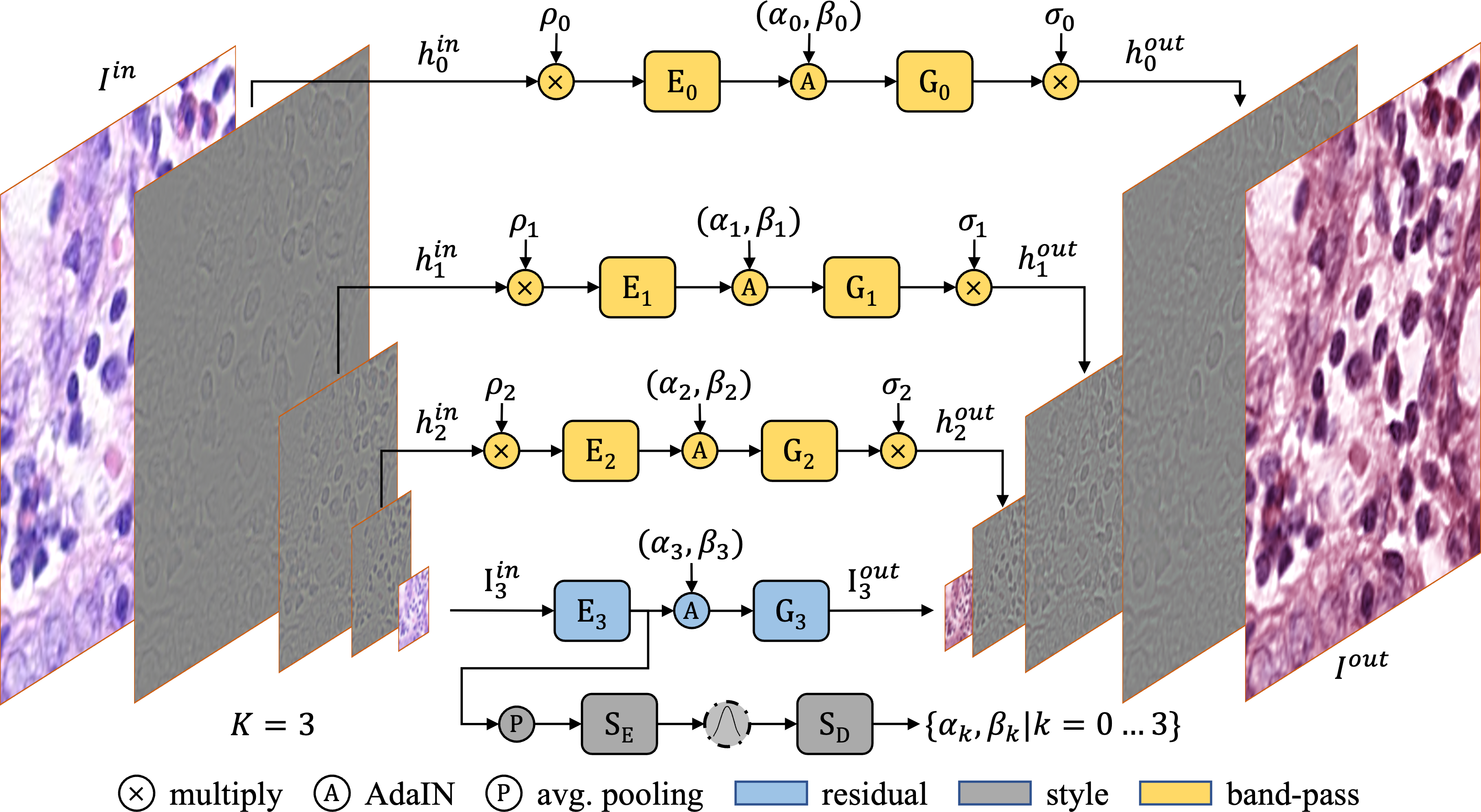}
\caption{
The G-SAN architecture for $K = 3$.
For any value of $K$, the architecture consists of three different pathways: residual, style, and band-pass (BP), each depicted with a different color in the figure.  
The residual pathway, shown in blue, produces the style-transferred low-resolution residual image at the output. 
The style pathway, shown at the bottom in gray is a Style Mapping Network (SMN) that is responsible for encoding and decoding the stain information. 
Finally, the multiple BP pathways independently produce the band-pass images at increasingly higher resolutions in the output LP pyramid.  
By allocating the computation-intensive operations only to the residual pathway and using only light-weight convolutional modules in the BP pathways, G-SAN avoids heavy convolutions at higher resolutions.  
Note that in the SMN, both the encoder and the decoder are implemented only with MLP layers, and the random resampling of latent stain vectors occurs only in the identity reconstruction mode during training.
}
\label{fig:architecture}
\end{figure*}

\begin{table*}[t]
\caption{
Convolutional layer specifications of the G-SAN generator.
All \texttt{conv2D} modules use \texttt{kernel\_size=3}.
}
\label{tbl:arch}
\centering
\begin{tabular}{@{}lll@{}}
\toprule
 & Encoder $E$ & Generator $G$ \\ \midrule
\multirow{2}{*}{Level $k = 0, ..., K-1$} & \texttt{conv2D}$(3, k \times 16)$, \texttt{LeakyReLU} & \texttt{LeakyReLU}, \texttt{conv2D}$(k\times32, k\times16)$ \\
 & \texttt{conv2D}$(k \times 16, k \times 32)$ & \texttt{LeakyReLU}, \texttt{conv2D}$(k \times 16, 3)$ \\ \midrule
\multirow{5}{*}{Level $K$} & \texttt{conv2D}$(3, 16)$, \texttt{LeakyReLU} & \texttt{LeakyReLU}, \texttt{ResBlock}$(256, 128, \texttt{LayerNorm}, \texttt{LeakyReLU})$ \\
 & \texttt{conv2D}$(16, 64)$, \texttt{LeakyReLU} & \texttt{ResBlock}$(128, 64, \texttt{LayerNorm}, \texttt{LeakyReLU})$ \\
 & \texttt{ResBlock}$(64, 128, \texttt{LayerNorm}, \texttt{LeakyReLU})$ & \texttt{conv2D}$(64, 16)$, \texttt{LeakyReLU} \\
 & \texttt{ResBlock}$(128, 256, \texttt{LayerNorm}, \texttt{LeakyReLU})$ & \texttt{conv2D}$(16, 3)$ \\
 & \texttt{conv2D}$(256, 256)$ & \\ \bottomrule
\end{tabular}
\end{table*}

The network architecture of G-SAN for image-to-image translation is shown in \cref{fig:architecture}.  
The input to G-SAN is the LP representation of the input image and, correspondingly, the output of G-SAN is also an LP representation from which the output image can be reconstructed.  
The generator architecture can be broken down into three pathways: residual, style, and band-pass (BP).  
By optimizing each pathway to produce a component of the output LP representation, we are able to achieve structure-preserving stain transfer with great computational efficiency.

Starting with the residual pathway, shown in blue, it is implemented as an encoder-generator pair and it works in conjunction with the style mapping pathway, shown in gray, that is implemented as an autoencoder.  
Let $\mI^{in}$ and $\mI^{out}$ denote the input image and the output stain-transferred image, respectively.  
The residual pathway, whose parameters are presented in \cref{tbl:arch}, is responsible for producing the stain-transferred low-resolution residual image $\mI_K^{out}$.
First, the encoder $E_K$ encodes $\mI_K^{in}$, the input LP image at level $K$, into a deep encoding $\vz_K^{in}$.
Subsequently, the stain vector of the input image $\vz_s^{in}$ is extracted by the style encoder $S_E$ from the deep encoding $\vz_K^{in}$. 
To achieve stain transfer, the target low-level deep encoding $\vz_K^{out}$ is produced by applying AdaIN on $\vz_K^{in}$, with the AdaIN parameters $(\text{mean}, \text{std}) = (\alpha_K, \beta_K)$ supplied by the style decoder $S_D$, shown in gray at the bottom of \cref{fig:architecture}.
Finally, the output image $\mI_K^{out}$ is generated from $\vz_K^{out}$ by the low-level generator $G_K$.

The task of the BP pathways is to adjust the input band-pass images for stain transfer at levels $k = 0$ to $K - 1$.
At level $k$, the input to the encoder $E_k$ is $\vh_k^{in}$, the input LP image at level $k$.
Similar to what is done in the residual pathway, the input is mapped to a deep encoding and subsequently transformed using AdaIN, where the target normalization parameters $(\alpha_k, \beta_k)$ are supplied by the style decoder $S_D$. 
The resulting target deep encoding is then mapped to the target LP representation $\vh_k^{out}$.
Compared to the low-level pathway, which consists of computationally heavy residual blocks, the BP pathways are implemented with light-weight convolutional modules using decreasing numbers of filters as resolution increases as shown in \cref{tbl:arch}.

It is important to note that we scale both the input and output of the BP pathway at level $k$ with \textit{non-learnable} scalars, $\rho_k$ and $\sigma_k$, respectively.
This is necessary due to the fact that, since the band-pass images capture only the high-frequency details, they generally have zero mean and significantly smaller standard deviations than the residual image.

Therefore, by applying the scale factors, we benefit the learning of the band-pass pathways by ensuring the dynamic range of the input image to $E_k$ and the output image by $G_k$ to be close to $(-1, 1)$, similar to what it would be for the residual images.
In our implementation, we choose the value of $\sigma_k$ to be the precalculated absolute max value of $\vh_k$ averaged from all training images and set $\rho_k = 1 / \sigma_k$.
Additionally, we found that making the scaling factors non-learnable can further stabilize the initial phase of training, where the quality of the generated BP images can be particularly sensitive.

Lastly, once we have obtained all the stain-transferred band-pass images and the residual image, the target image can be produced by applying the backward recurrence in \cref{eq:lp-recon}.

\subsection{Disentangling Morphology from Stain}
\label{subsec:disentangle}
To enable structure-preserving style transfers between arbitrary stains, the stain representation must first be fully disentangled from the underlying morphology representation. 
With LP representations, while the stain information is the most prevalent in the low-res residual image $\mI_K$, it is also evident albeit more weakly in the band-pass images $\vh_k$. 
As mentioned previously in \cref{subsec:lp}, this phenomenon is clearly visible in the histograms plotted in \cref{fig:histogram}. 
Therefore, it is necessary to achieve morphology-stain disentanglement in all levels of the LP representation, which has not been carried out in previous LP-based image-to-image translation networks, \eg \cite{liang2021high,lin2021drafting}.

In G-SAN, we assume that the stain information can be fully captured by the channel normalization parameters of the convolutional features. 
Therefore, we use instance normalization (IN) as the model bias that removes any stain-related information from the deep encodings in the pathways and the resulting normalized encodings represent only the morphology.  
Subsequently, by applying the AdaIN parameters $(\alpha, \beta)$ to the purely morphological encoding, we can transfer the target stain to the encoding.
In G-SAN, the set of $(\alpha_k, \beta_k)$ parameters for a target stain is provided by the style decoder $S_D$ in the Style Mapping Network.

\begin{figure*}[t]
\centering
\includegraphics[width=0.9\textwidth]{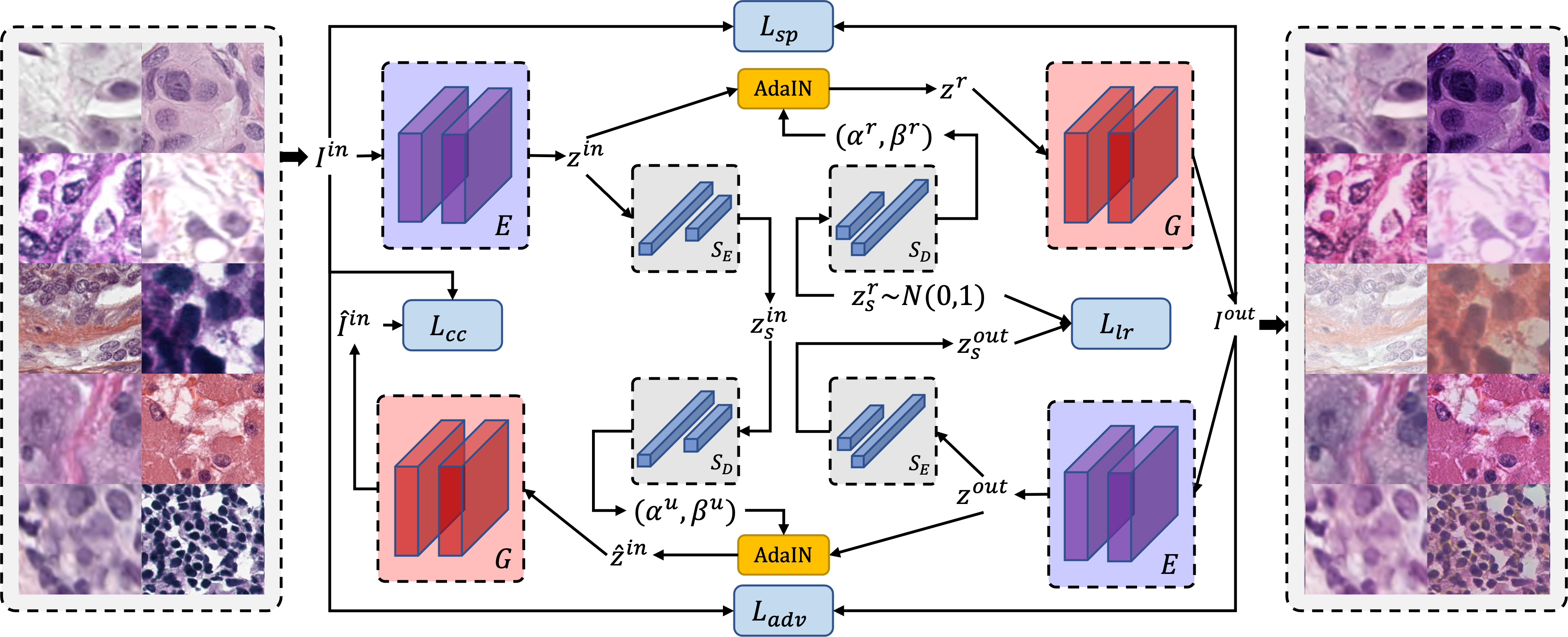}
\caption{
This figure presents an overview of the cyclic reconstruction mode (Mode B) of the training procedure for G-SAN. 
In the forward direction, given an input image $\mI^{in}$, the encoding process produces a deep  encoding $\vz^{in}$ along with its stain encoding $\vz_{s}^{in}$.
Subsequently, the generative process combines $\vz^{in}$ with a noise stain encoding $\vz_{s}^{r}$ via AdaIN to produce a stain-augmented version of the input image, $\mI^{out}$.  
And in the reverse direction, the deep code $\vz^{out}$ is first extracted from $\mI^{out}$, then  combined with the original stain encoding $\vz_{s}^{in}$ via AdaIN, and finally passed to $G$ to produce the cyclic reconstruction $\hat{\mI_k^{in}}$.
}
\label{fig:training}
\end{figure*}

\subsection{Handling Multiple Resolutions}
The LP-based image representation is recursive in the sense that the LP representation $L(\mI_k)$ of the image $\mI_k$ at level $k$ can be decomposed into a band-pass image $\vh_k$ and the LP representation $L(\mI_{k+1})$ of the image one level below.  
Owing to that recursive nature, a single stain transfer network trained to process the LP representations in the highest resolution can be readily used for input images with lower resolutions. 
This makes our framework particularly versatile since the pathology images are often recorded at different resolutions for different tasks. 
For example, for nucleus segmentation the images are often used at $40\times$ magnification level and for tissue phenotyping at $20\times$.  
If we train the LP-based generator to produce images at $40\times$, the same network can be readily used for $20\times$ images just by ignoring the BP pathway at $k = 0$ and using instead the output image reconstructed at $k = 1$. 
Along the same lines, $10\times$ images can be processed and reconstructed at $k = 2$ using the G-SAN generator trained with images at $40\times$. 
What that implies is that, with no additional training and no extra architectural elements, our LP-based model can be considered to be generalized across a range of image resolutions.

During the training of G-SAN, we leverage the concept of deep supervision and calculate the image reconstruction loss at each LP level. 
Similarly, we also employ a multi-resolution discriminator that consists of identical purely convolutional networks at each level to encourage output images at all levels to be realistic.
The next subsection presents further details regarding these aspects of G-SAN.

\subsection{The Training Procedure and the Losses}
\label{subsec:training}

For brevity (but without compromising essential details), the presentation in this section is in terms of relatively high-level abstractions. 
We will therefore ignore the specific architectural details related to the Laplacian Pyramid.
Given the network components -- $E$ as the encoder, $G$ as the generator, $S$ as the SMN and $D$ as the discriminator -- the encoding process for an input image $\mI^{in}$ can be written as:
\begin{equation}
\vz^{in} = E(\mI^{in}) ~~ \text{and} ~~ \vz_{s}^{in} = S_E(\vz^{in}).
\end{equation}
The generative process, on the other hand, can happen in one of the two modes: \textbf{Mode A} -- the identity reconstruction mode; and \textbf{Mode B} -- the cyclic reconstruction mode (\cref{fig:training}). 
In Mode A, the identity reconstruction $\tilde{\mI^{in}}$ can be written as:
\begin{equation}
\tilde{\vz^{in}} = \text{AdaIN}(\vz^{in}, S_D(\tilde{\vz_{s}^{in}})) ~~ \text{and} ~~ \tilde{\mI^{in}} = G(\tilde{\vz^{in}}),
\end{equation}
where $\tilde{\vz_{s}^{in}}$ is a resampled version of $\vz_{s}^{in}$ obtained through the reparameterization trick for VAE (Variational Autoencoder). 
The losses calculated in the identity reconstruction mode are as follows: \\ 
\textbf{Identity Reconstruction Loss} ensures the learned encodings
$\vz$ and $\vz_s$ to be representative enough to recover the original input image. 
This image reconstruction loss is a weighted sum of losses at all levels of the image output:
\begin{equation}
\calL_{id}(\mI^{in}, \tilde{\mI^{in}}) = \mathbb{E}_{\mI^{in}} \left[ \sum_k m_k \left\| \mI_k^{in} - \tilde{\mI_k^{in}} \right\|_1 \right].
\end{equation}
\textbf{VAE Loss} encourages the latent stain vectors from the images actually recorded to conform to a prior Gaussian distribution to facilitate stochastic sampling at test time.  
It is calculated through the KL-divergence:
\begin{equation}
\calL_{vae}(\vz_{s}^{in}) = \mathbb{E}_{\vz_{s}^{in}} \left[ D_{\text{KL}}(\vz_{s}^{in} || N(0,1)) \right],
\end{equation}
where $D_{\text{KL}}(p || q) = - \int p(z) \log \frac{p(z)}{q(z)} \text{d}z$.

In Mode B, the random augmentation $\mI^{out}$ and the cyclic
reconstruction $\hat{\mI^{in}}$ are given as:
\begin{equation}
\mI^{out} = G(\vz^r) = G(\text{AdaIN}(\vz^{in}, S_D(\vz_{s}^{r}))), 
\end{equation}
\begin{equation}
\text{and} ~~ \hat{\mI^{in}} = G(\text{AdaIN}(\vz^{out}, S_D(\vz_{s}^{in}))),
\end{equation}
where $\vz_{s}^{r}$ denotes a randomly sampled stain vector.  
The relevant losses are: \\ 
\textbf{Cross-Cycle Consistency Loss} constrains the
cross-cycle-reconstructed version to be consistent with the original
input image in multiple resolutions:
\begin{equation}
\calL_{cc}(\mI^{in}, \hat{\mI^{in}}) 
= \mathbb{E}_{\mI^{in}} \left[ \sum_k m_k \left\| \mI_k^{in} - \hat{\mI_k^{in}} \right\|_1 \right].
\end{equation}
\textbf{Structure-Preserving Loss} is an adaptation of the perceptual
loss introduced in \cite{johnson2016perceptual} -- the instance
normalization function is applied on each set of features extracted by
$\phi(\cdot)$ at level $i$:
\begin{equation}
\calL_{sp}(\mI^{in}, \mI^{out}) 
= \mathbb{E}_{\mI^{in}} \left[ \sum_i^N \frac{1}{w_i h_i d_i} \left\| \text{IN}( \phi_i(\mI^{in}) ) - \text{IN}( \phi_i(\mI^{out}) ) \right\|_F^2 \right],
\end{equation}
where $\|\cdot\|_F$ denotes the Frobenius norm, and $w$, $h$ and $d$ represent the width, height and depth of the feature space.
As shown in \cite{huang2018multimodal}, applying instance normalization makes the loss more domain-invariant.
This is particularly important in our case since it penalizes undesirable alterations to cell morphology by stain transformation. \\
\textbf{Latent Regression Loss} helps prevent mode collapse by encouraging a reversible mapping between the stain latent space and the image space:
\begin{equation}
\calL_{lr}(\vz_{s}^{r}, \vz_{s}^{out}) = \mathbb{E}_{\vz_{s}^{r} \sim N(0, 1)}  \left[ \left\| \vz_{s}^{r} - \vz_{s}^{out} \right\|_1 \right].
\end{equation}
\textbf{Mode Seeking Loss} encourages the randomly generated samples to be more diverse by minimizing the following ratio:
\begin{equation}
\calL_{ms}(\vz_{s}^{r_1}, \vz_{s}^{r_2}) = \mathbb{E}_{\vz_{s}^{r1}, \vz_{s}^{r2} \sim N(0, 1)} \left[ \frac{\left\| \vz_{s}^{r1} - \vz_{s}^{r2} \right\|_1}{\left\| \mI^{r1} - \mI^{r2} \right\|_1 + \epsilon} \right],
\end{equation}
where $\epsilon$ is a small stabilizing constant. \\
\textbf{Adversarial Loss} encourages the randomly stained images $\mI^{out}$ to be indistinguishable from the set of cell images actually recorded, in terms of both stain and morphology in multiple resolutions.  
The loss takes the form of least squares
\cite{mao2017least}:
\begin{equation}
\begin{aligned}
\calL_{adv}(E, G, D) &= \frac{1}{2} \mathbb{E}_{\mI^{out}} \left[  \sum_k D_k(\mI^{out}_k)^2 \right] \\
&+ \frac{1}{2} \mathbb{E}_{\mI^{in}} \left[ \sum_k \left( 1 - D_k(\mI^{in}_k) \right) ^2 \right].
\end{aligned}
\end{equation}

Finally, the combined min-max optimization objective for G-SAN from the two modes, Mode A and Mode B, can be written as:
\begin{equation}
\begin{aligned}
E^*, G^* =& \arg \fdlmax{E,G} \fdlmin{D} \calL_{adv} + \lambda_{id} \calL_{id} + \lambda_{vae} \calL_{vae} \\
&+ \lambda_{cc} \calL_{cc} + \lambda_{sp} \calL_{sp} + \lambda_{lr} \calL_{lr} + \lambda_{ms} \calL_{ms},
\end{aligned}
\label{eq:total-loss}
\end{equation}
where the $\lambda$s are tunable hyperparameters.

\section{Experimental Results}
\label{sec:exp}

\begin{figure*}[t]
\centering
\includegraphics[width=0.99\textwidth]{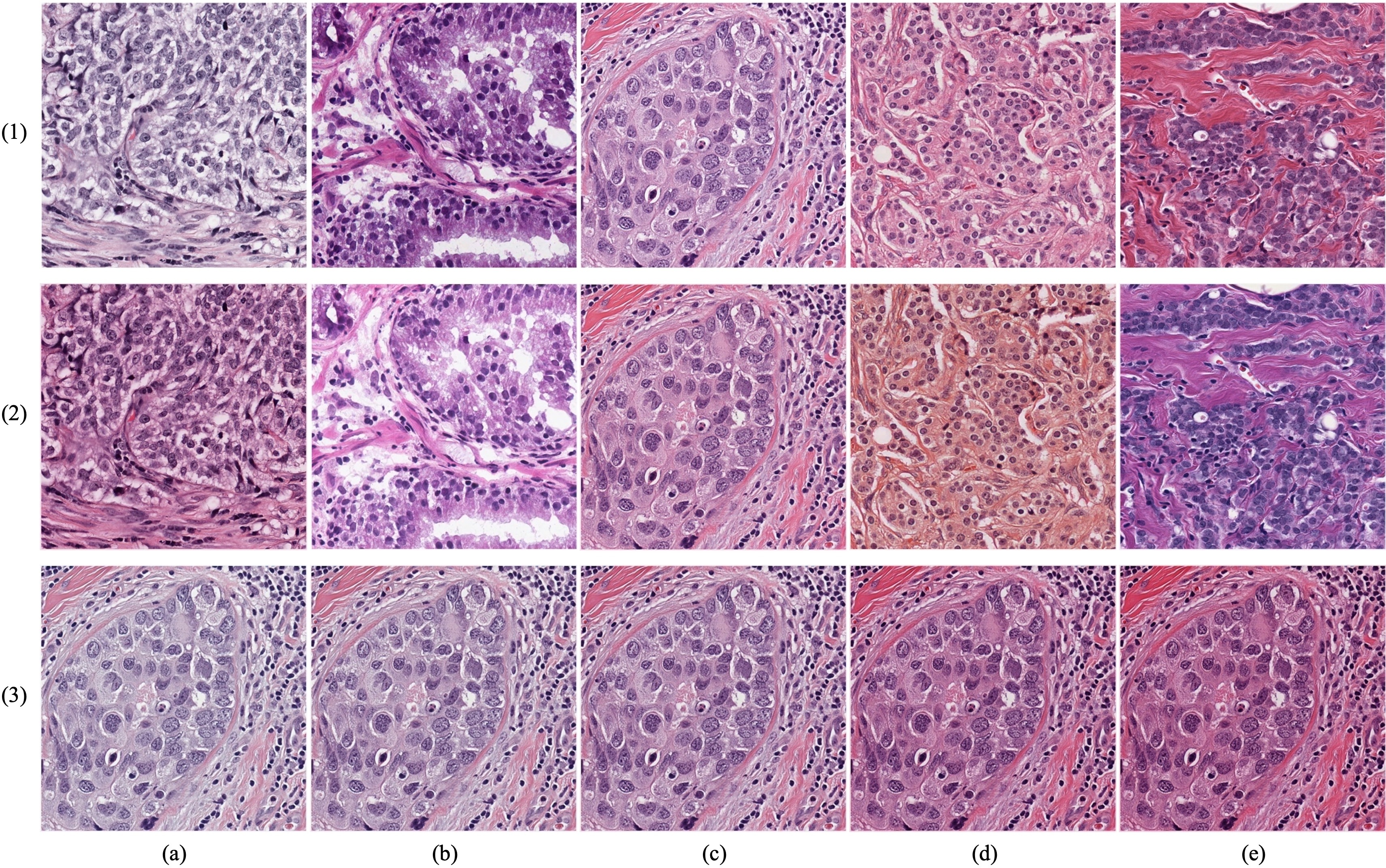}
\caption{Row (1): images from \cite{kumar2019multi}; Row (2): input images augmented by G-SAN; Row (3): interpolation results by mixing the morphology from image (1c) with the stains obtained through linearly interpolating between the stain vectors from image (1a) and (1e).}
\label{fig:random}
\end{figure*}

\begin{figure*}[t]
\centering
\includegraphics[width=0.9\textwidth]{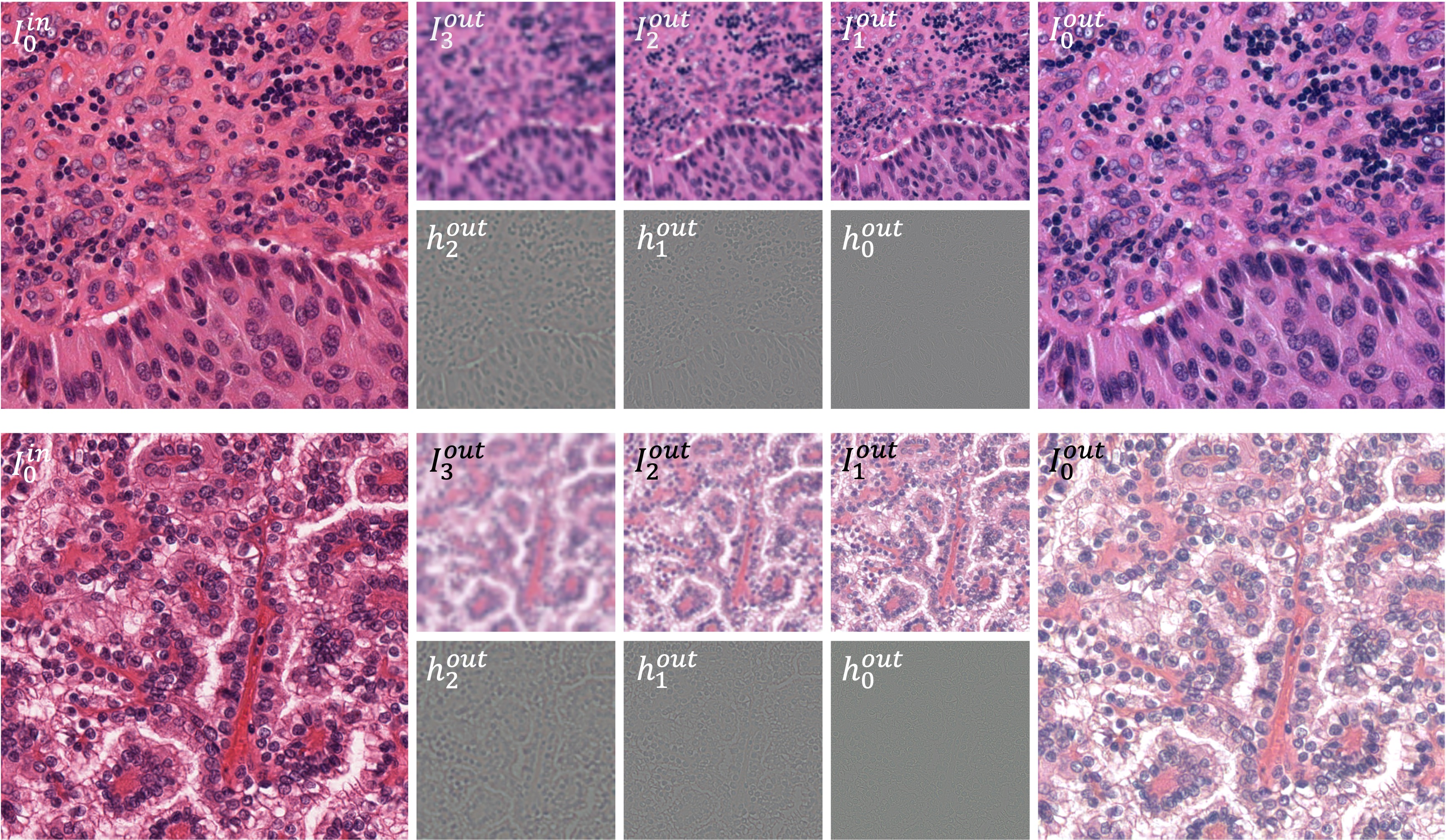}
\caption{
Dissecting the G-SAN augmented images.
For the stain-augmented version of an input image $\mI_{k=0}^{in}$ at $40\times$, G-SAN produces both the Gaussian Pyramid (GP), $G(\mI^{out}) = [\mI_1^{out}, \mI_2^{out}, \mI_3^{out}]$, as well as the Laplacian Pyramid (LP), $L(\mI^{out}) = [\vh_0^{out}, \vh_1^{out}, \vh_2^{out}, \mI_3^{out}]$ that is used to construct the GP.
Note that in the figure, the $\mI_{k = 2,3}^{out}$ and $\vh_{k = 0,2}^{out}$ images have been resized to fit the display grid.
Please zoom in to see the structures in the reduced-size images.
}
\label{fig:gsan_laplacian}
\end{figure*}

\begin{figure}[t]
\centering
\includegraphics[width=0.49\textwidth]{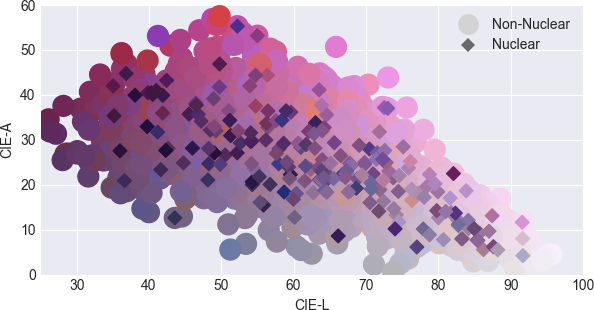}
\caption{A scatter plot of the most dominant colors in the cell images produced by G-SAN. 
Through the stochastic sampling of a normal distribution in the stain latent space as learned by the SMN, a diverse yet realistic distribution of stain appearances can be achieved with regard to both hue and lightness. 
Note that the nuclear and the non-nuclear regions were separated using ground-truth masks and their most dominant colors were extracted using the median-cut algorithm reported in \cite{fengsp}. 
The axes correspond to the non-nuclear colors. 
Only a subset of the nuclear points is shown for a less cluttered visualization.
}
\label{fig:latent}
\end{figure}

The training dataset for G-SAN consists of patches extracted from 573 WSIs downloaded from the TCGA program \cite{tcga}. 
The selection of WSIs is carefully curated to maximize the diversity in terms of both the H\&E stain appearance and cell morphology.  
More specifically, with each WSI representing a unique pair of (tissue site, laboratory ID), there are 33 tissue sites from around 200 laboratories included in our training data\footnote{A comprehensive superset of the WSI origins can be found at \cite{tcgatss}.}.
In total, we extracted 348k patches of size $512 \times 512$ at $40\times$ magnification. 
We trained G-SAN for 60k iterations using the ADAM optimizer with a linear-decay learning-rate scheduler with the initial learning rate set to $1e^{-4}$.
Training took about 9 hours with an AMD 5800X 8-core CPU with 32G RAM and a Nvidia RTX3090 GPU with 24G memory.
The hyperparameters in \cref{eq:total-loss} are set as $\lambda_{id} = 1$, $\lambda_{vae} = 0.01$, $\lambda_{cc} = 10$, $\lambda_{sp} = 0.5$, $\lambda_{lr} = 10$, and $\lambda_{ms} = 0.02$.
See \cref{subsec:lambda} for how we arrived at these values for the hyperparameters.

In the rest of this section, we first provide a qualitative analysis of G-SAN augmentations, followed by quantitative analyses through two common downstream tasks: patch classification at $20\times$ magnification and nucleus segmentation at $40\times$.
All experimental results were obtained with a single G-SAN model where $K = 3$. 

We denote this model as $\text{G-SAN}_{K=3}$ and it is used for both downstream tasks in our quantitative analysis.
The notation ``$\text{G-SAN}_{K=3}$ @ $k = 0$'' indicates that the image inputs and outputs of G-SAN are given and taken at pyramid level $k = 0$ (\ie at $40\times$ magnification), while $k = 1$ corresponds to $20\times$.
Furthermore, we provide a timing analysis comparing several commonly used stain transfer and stain augmentation tools to G-SAN.
Lastly, we offer insights into some of the design choices in G-SAN through ablation studies.

\subsection{Qualitative Analysis}
In rows (1) and (2) of \cref{fig:random}, we first showcase the G-SAN-augmented results -- note how G-SAN is able to augment cell images that are diverse in both cell morphology and stain colors.  
In row (3), we performed linear interpolations between two stain encodings extracted from two stain-reference images and combined the interpolated stain codes with the morphology code extracted from a morphology-reference image. 
The fact that applying the interpolated stains resulted in smooth changes in the images shown in the last row illustrates that the latent space is generally smooth, which is a desirable property if it is to lend itself to stochastic sampling.
Subsequently in \cref{fig:gsan_laplacian}, we showcase the multi-resolution stain-augmented outputs by G-SAN, along with the generated band-pass images.
Especially note how realistic the generated band-pass images are when compared to those from the LPs of real images in \cref{fig:histogram}.
Lastly, to visually demonstrate the range of stain appearances covered by the latent space, \cref{fig:latent} is a scatter plot of the most dominant colors from the cell images that are produced by G-SAN.

\subsection{Downstream Task I: Patch Classification}
\label{subsec:c17}

\begin{figure}[t]
\centering
\includegraphics[width=0.49\textwidth]{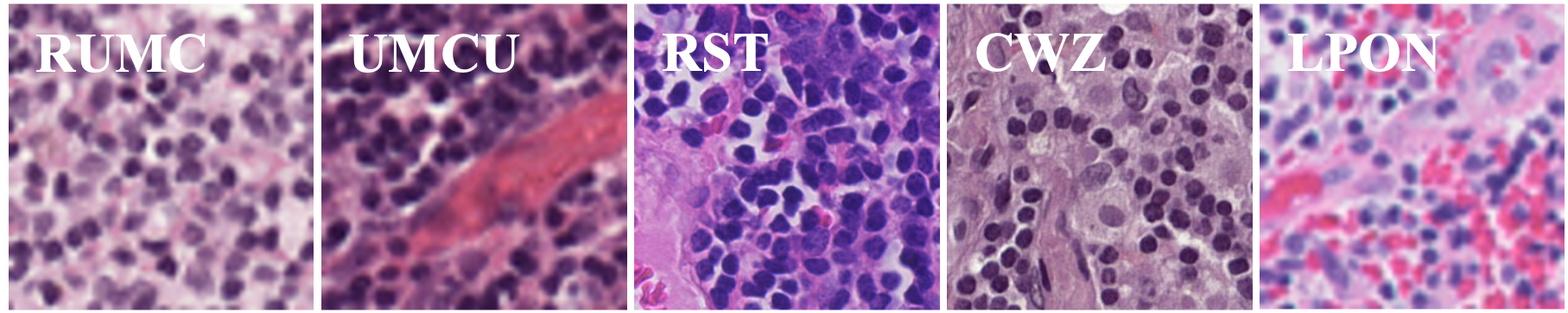}
\caption{Example patches from the five medical centers in the CAMELYON17 dataset.}
\label{fig:c17}
\end{figure}

\begin{figure*}[t]
\centering
\includegraphics[width=0.99\textwidth]{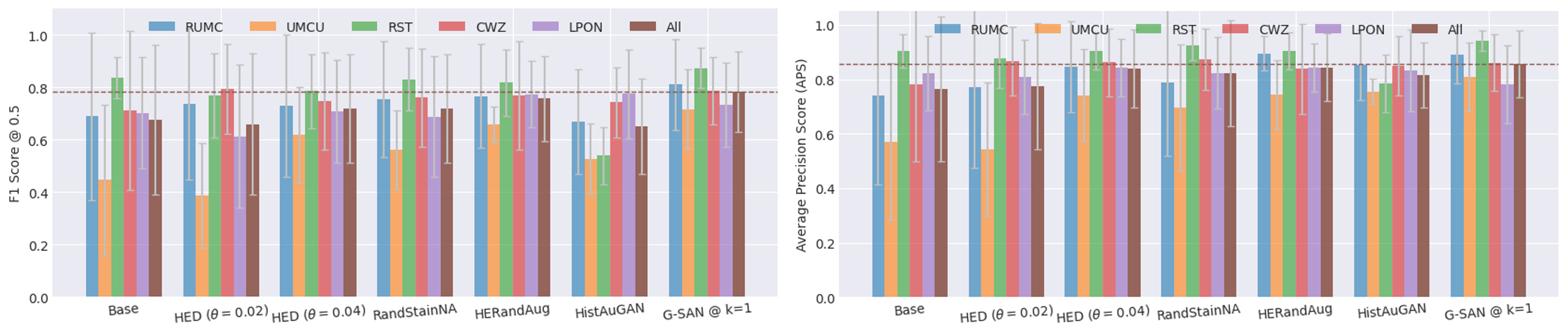}
\caption{
F1 scores and Average Precision Scores (APS) of the tumor class for our 5-fold cross-validated patch classification experiment on the CAMELYON17 dataset.
For the G-SAN results shown, the input images and the outputs produced are for the pyramid level $k = 1$ (\ie at $20\times$ magnification).
}
\label{fig:c17_results}
\end{figure*}

For the first quantitative assessment, we choose the downstream task of patch classification of breast cancer metastases using the CAMELYON17 dataset \cite{bandi2018detection}.
We used the semantically labeled subset, comprising 50 WSIs of histological lymph node sections with metastasis regions labeled at pixel level.
It is important to note that the WSIs were made at 5 different medical centers with 10 WSIs per center.
On account of the differences in the staining procedures used and also the differences in the imaging equipment across the 5 medical centers, there exist significant stain variations among the resulting images. Example patches demonstrating the varying stains are shown in \cref{fig:c17}.
We preprocessed the tissue regions in the WSIs with patches at $20\times$ magnification level, resulting in a total of 210k non-overlapping patches of size $256 \times 256$.
We followed the same practice as described in \cite{wagner2021structure} for label assignment: if the tumor masked region exceeds 1\% in a patch, the patch is labeled positive.

In our 5-fold cross-validated experiment, we perform training and validation of our classification network only on patches from a single medical center in each fold.
This is to simulate the practical scenario in which the available labeled training data is scarce and has limited stain variation.  
Patches from the other four centers are therefore out-of-domain in terms of the stain and used as testing data. 
Additionally, note that positive and negative patches are drawn with equal probabilities during training and validation.
The results obtained with the different stain augmentation approaches are shown in \cref{fig:c17_results}.
In addition to the simple \textit{HED Jitter} augmentations, we also compare G-SAN to the state-of-the-art in non-learning based stain augmentation frameworks, such as HERandAugment \cite{faryna2021tailoring} and RandStainNA \cite{shen2022randstainna}.
For both HistAuGAN \cite{wagner2021structure}
\footnote{For HistAuGAN, we used the pretrained weights provided by its authors on patches at $40\times$ from the five domains of the CAMELYON17 dataset.
For stain augmentation, we used a randomly interpolated domain as the target domain for each image.}
and G-SAN, the stain vectors were randomly drawn from a normal distribution. 
\new{In our dataloader, stain augmentation was applied to every image loaded for training. 
Stain augmentation was also applied to the images loaded for validation to prevent statistically biased evaluations of our models due to the limited stain appearances in the validation data.}
\new{Additionally, we believe that a stain augmentation method is worthy of merit only if it can also diversify the validation stain distribution such that the validation score better correlates with the true generalizability of a model.}

From the results in \cref{fig:c17_results}, we can first confirm the domain gaps among the images taken at different medical centers, as the scores by the baseline method (\ie without stain augmentation) vary greatly across the folds.
Such domain gaps can be effectively reduced by applying stain augmentations.  
Additionally, among the stain augmentation methods, it can be observed that augmentations by G-SAN are the most effective, as they provide the greatest boosts in both the overall F1 score (15.7\%) and the overall Average Precision Score (12.1\%) compared to the baseline.
Given that the second best performer, HERandAugment \cite{faryna2021tailoring}, produces unrealistic stain appearances by design, the superior performance by G-SAN just shows that augmenting cell images beyond the distribution of naturally occurring stain appearances may not be the best strategy.
Additionally, the poor performance by HistAuGAN could be attributed to its inflexibility towards multi-resolution, given that it was trained on images at $40\times$ magnification.
\new{Last but not least, it is worth mentioning that, as it cannot be avoided, sometimes the stain distribution of the unaltered training data can overlap better with the test stain distribution.
However, in most cases as shown in our experiments, using any form of stain augmentation will provide a boost in performance.}

\subsection{Downstream Task II: Nucleus Segmentation}
\label{subsec:nucseg}

\begin{figure*}[t]
\centering
\includegraphics[width=0.99\textwidth]{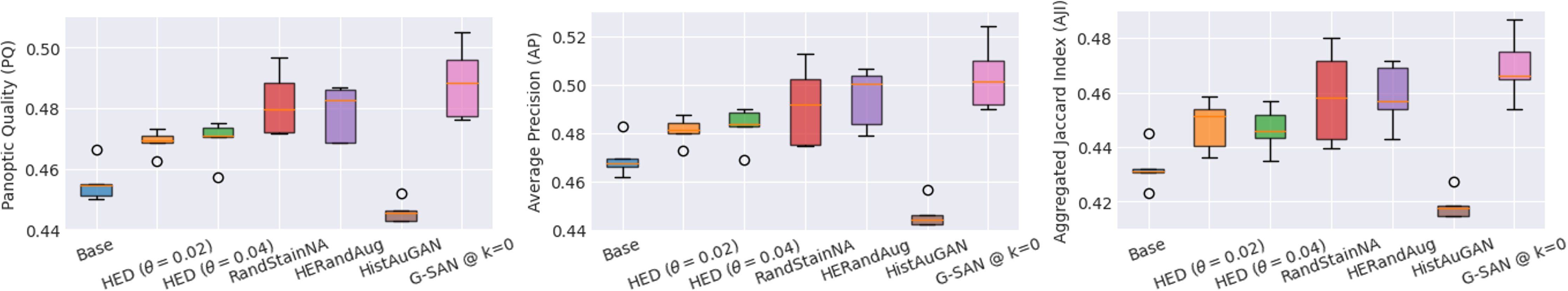}
\caption{
Panoptic Quality (PQ), Average Precision (AP) and Aggregated Jaccard Index (AJI) scores of the 5-fold nucleus segmentation experiment.
\new{The images used were collected from the following publicly available datasets: MoNuSeg \cite{kumar2019multi}, CPM15, CPM17 \cite{vu2019methods}, CryoNuSeg \cite{mahbod2021cryonuseg}, MoNuSAC \cite{verma2020multi}, TNBC \cite{naylor2018segmentation}, and CoNSeP \cite{graham2019hover}. 
More details about each dataset can be found in \cref{tbl:nucseg_data}.}
For the G-SAN results shown, the input images and the outputs produced are for the pyramid level $k = 0$ (\ie at $40\times$ magnification).
}
\label{fig:nucseg_results}
\end{figure*}

\begin{table*}[t]
\centering
\caption{
\new{Full details on the datasets used in our nucleus segmentation experiment.}
}
\label{tbl:nucseg_data}
\begin{tabular}{@{}cp{0.4\linewidth}cc@{}}
\toprule
Dataset & \multicolumn{1}{c}{Tissue Site} & Image Size & Quantity \\ \midrule
MoNuSeg \cite{kumar2019multi} & Kidney, Lung, Colon, Breast, Bladder, Prostate, Brain & $1000 \times 1000$ & 44 \\
CPM \cite{vu2019methods} & Lung, Head and Neck, Brain & $[439, 1032] \times [392, 888]$ & 79 \\
CryoNuSeg \cite{mahbod2021cryonuseg} & Adrenal Gland, Larynx, Lymph Node, Mediastinum, Pancreas, Pleura, Skin, Testis, Thymus, Thyroid Gland & $512 \times 512$ & 30 \\
MoNuSAC \cite{verma2020multi} & Lung, Prostate, Kidney, Breast & $[35, 2162] \times [33, 2500]$ & 294 \\
TNBC \cite{naylor2018segmentation} & Breast, Brain & $512 \times 512$ & 68 \\
CoNSeP \cite{graham2019hover} & Colon & $1000 \times 1000$ & 41 \\ \bottomrule
\end{tabular}
\end{table*}

We have also evaluated the performance improvements made possible by the augmentations generated by G-SAN on the downstream task of nuclear instance segmentation.
Nuclear instance segmentation is challenging due to high morphological and textural heterogeneity of the nuclei as well as their small sizes. 
What that implies is that any stain augmentation framework must be highly structure preserving in order to be useful.  
In our experiments with nuclear segmentation, we used a straightforward gradient-flow based CNN model inspired by \cite{graham2019hover,stringer2021cellpose}.
To quantitatively measure the instance segmentation quality, we use the Panoptic Quality (PQ) as defined in \cite{graham2019hover}, the Average Precision (AP) in \cite{stringer2021cellpose} as well as the Aggregated Jaccard Index (AJI) in \cite{kumar2019multi}.

In light of the limited quantity of the available nucleus groundtruth, we evaluated nucleus segmentations with 5-fold cross-validation as explained in what follows.
\new{In total, we curated 556 images at $40\times$ magnification with nucleus annotations from six publicly available datasets as tabulated in \cref{tbl:nucseg_data}.}
\new{Since each dataset covers a different set of organs, and the cell morphology varies considerably across organs, we cannot train a model on a single dataset and expect it to generalize well to the others.}
\new{As a result, we grouped images from all the dataset together and divided them into 5 folds.}
Images from one fold are used for training and validation, while images from the other four folds are used for testing.
\new{Given the scarcity of nucleus annotations, our cross-validation setup simulates the realistic scenario where the quantity of available labeled data for training and validation is on the same level as in most of the publicly accessible datasets as listed in \cref{tbl:nucseg_data}.}
\new{Moreover, complimentary to what was the case for the CAMELYON17 dataset we used for patch classification, each fold here represents a wide range of organs and covers a diverse set of stain appearances.}
\new{With this cross-validation setup, we hope to demonstrate that G-SAN can benefit the training of generalized models for nucleus segmentation across organs, which is in the interest of researchers \cite{stringer2021cellpose}.}

From the test scores plotted in \cref{fig:nucseg_results}, we can again observe that G-SAN offers the largest average improvement over the baseline (\ie without stain augmentation) in terms of all three metrics: 7.3\% in PQ, 7.2\% in AP and 8.5\% in AJI.
Regarding the performance of HistAuGAN, while a cursory examination of the stain augmentations generated by the network may cause one to think that they are of high quality, the reality is that the augmentations are not structure-preserving and therefore the algorithm comes up short from the standpoint of producing good segmentations.
\new{This shortcoming of HistAuGAN could be attributed to the significant heterogeneity in tissue morphology across organs, coupled with the fact that it was exclusively trained on breast cancer images from the CAMELYON17 dataset \cite{wagner2021structure}.}

\subsection{Timing Analysis}
\label{subsec:timing}

In \cref{tbl:timing}, we tabulate the average time per image needed for stain augmentation for a range of image sizes.
We compare the run times of G-SAN against CPU-based implementations of the SOTA in stain separation (\ie Macenko \cite{macenko2009method} and Vahadane \cite{vahadane2016structure}), as well as the competing stain augmentation methods used previously in the downstream tasks.
\new{With the stain separation methods, while we recognize that their efficiency can be optimized with prior knowledge of the data, we do not consider any application-specific or data-specific factors in our timing measurements for the sake of simplicity, especially given that the availability of such information is not guaranteed in practice.}
The experiments were conducted on the same machine with an AMD 5800X 8-core CPU and a Nvidia RTX3090 GPU. 
The run times are averaged over 1000 iterations.
Compared to all other stain transfer and stain augmentation methods, G-SAN is more scalable with increasing image dimensions.
Given input images of size $2048^2$, performing stain transfer using G-SAN at level 0 only requires up to 44\% of time needed by the fastest CPU-based multi-threaded stain separation or stain augmentation method.

\begin{table}[t]
\centering
\caption{
Seconds needed per image for stain transfer or stain augmentation using different methods. The best and the second best timings are denoted with \textbf{bold} fonts and $\dagger$, respectively.
}
\label{tbl:timing}
\begin{tabular}{lcccc}
\toprule
\multicolumn{1}{c}{Image Size} & $256^2$ & $512^2$ & $1024^2$ & $2048^2$ \\ \midrule
Macenko @ StainTools \cite{staintools} & 0.0199 & 0.0726 & 0.2754 & 1.1154 \\
Vahadane @ StainTools & 1.0191 & 1.0634 & 1.2243 & 1.9868 \\
Macenko @ TorchStain \cite{barbano2022torchstain} & 0.0076 & 0.0279 & 0.1063 & 0.5391 \\
HED Jitter \cite{tellez2019quantifying} & 0.0037\textsuperscript{$\dagger$} & 0.0141 & 0.0612 & 0.2664 \\
HERandAugment \cite{faryna2021tailoring} & 0.0090 & 0.0329 & 0.1279 & 0.5269 \\
RandStainNA \cite{shen2022randstainna} & \textbf{0.0024} & 0.0117 & 0.0433 & 0.1845 \\
HistAuGAN \cite{wagner2021structure} & 0.0171 & 0.0727 & 0.2946 & 1.2045 \\
$\text{G-SAN}_{K=3}$ @ $k=1$ & 0.0060 & 0.0113\textsuperscript{$\dagger$} & 0.0420\textsuperscript{$\dagger$} & 0.1664\textsuperscript{$\dagger$} \\
$\text{G-SAN}_{K=3}$ @ $k=0$ & 0.0049 & \textbf{0.0060} & \textbf{0.0209} & \textbf{0.0811} \\ \bottomrule
\end{tabular}
\end{table}

\section{Discussion}
\label{sec:discussion}

\subsection{Ablation Studies on the G-SAN Architecture}
\label{subsec:ablation}

\begin{table}[t]
\centering
\caption{
Ablation studies on several design choices in G-SAN using the nucleus segmentation experiment.
}
\label{tbl:ablation}
\begin{tabular}{@{}lccc@{}}
\toprule
\multicolumn{1}{c}{Average Score} & PQ & AP & AJI \\ \midrule
Base (No Stain Aug.) & 0.4553 & 0.4696 & 0.4325 \\
$\text{G-SAN}_{K=3}$ @ $k=0$ & \textbf{0.4885} & \textbf{0.5034} & \textbf{0.4693} \\
$\text{G-SAN}_{K=4}$ @ $k=0$ & 0.4812 & 0.4914 & 0.4615 \\
$\text{G-SAN}_{K=5}$ @ $k=0$ & 0.4737 & 0.4853 & 0.4565 \\
G-SAN w/ learnable scaling & 0.4834 & 0.4934 & 0.4642 \\
G-SAN w/o BP scaling & 0.4812 & 0.4942 & 0.4606 \\ \bottomrule
\end{tabular}
\end{table}

In this section, we conduct additional ablation studies on some of the most important design choices in G-SAN. 
We used the same nucleus segmentation experimental setup as in \cref{subsec:nucseg} and the results are tabulated in \cref{tbl:ablation}.
Regarding the choice of $K$, we specifically chose $K=3$ for our final model because as one can observe in \cref{fig:gsan_laplacian}, the residual image $\mI_{k=3}$ (\ie at $5\times$ if $\mI_{k=0}$ is at $40\times$) is the lowest resolution where the network can still accurately extract the H\&E stain information.
For any $k>3$, the nuclei become indistinct from the other morphological structures and therefore it is challenging to extract the correct Hematoxylin representation.
A direct consequence of this inability to extract correct stain representations is inadequate stain-morphology disentanglement.
In \cref{tbl:ablation}, the relatively poor performances of $\text{G-SAN}_{K=4,5}$ illustrate this effect.

Additionally, we conducted experiments on $\text{G-SAN}_{K=3}$ without scaling factors at the BP pathways, and with learnable scaling factors.
The results presented in \cref{tbl:ablation} demonstrate the importance of our proposed approach to BP scaling for competitive performance.
Our experiments showed that proper scaling of BP inputs and outputs can help prevent the appearance of visual artifacts in generated BP images, particularly during the initial stages of training.

\subsection{Determining the $\lambda$ Hyperparameters}
\label{subsec:lambda}

\begin{figure}[t]
\centering
\includegraphics[width=0.49\textwidth]{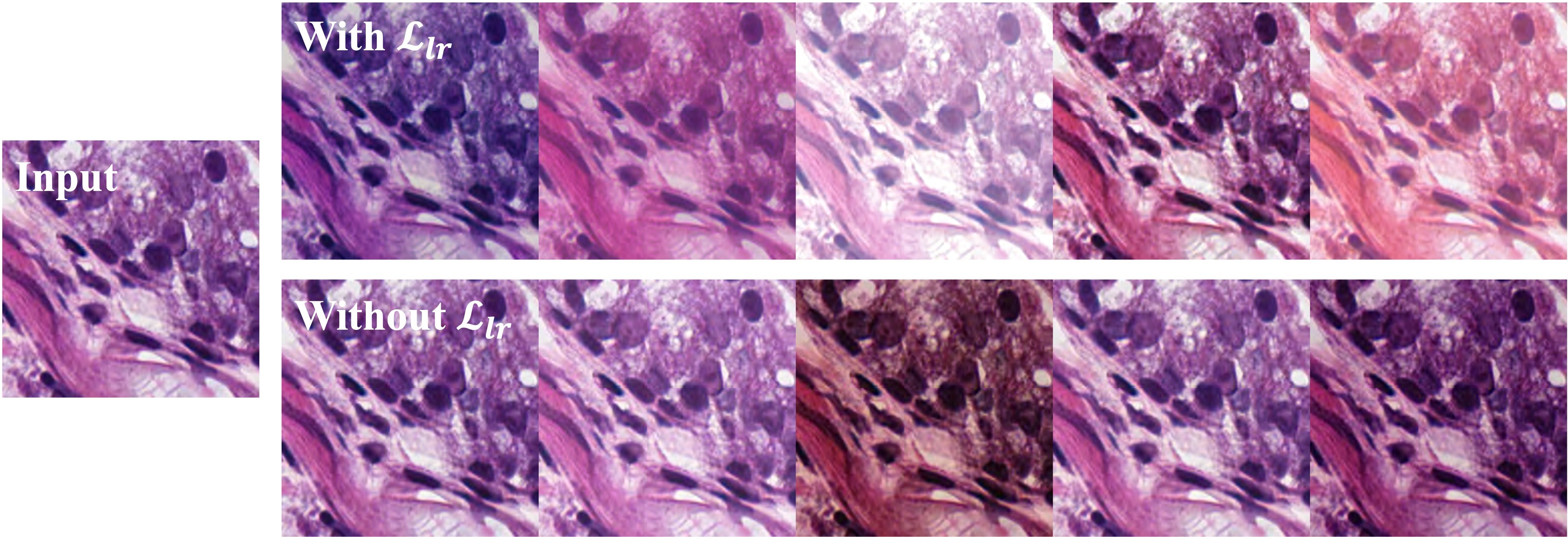}
\caption{
\new{Randomly stain-augmented patches by the G-SAN model trained with and without the latent regression loss $\calL_{lr}$.
Without $\calL_{lr}$, stain diversity of the augmented images is negatively impacted as the randomly sampled stain vectors can no longer contribute to the synthesized images as meaningfully.
}}
\label{fig:lr_ablation}
\end{figure}

This section outlines the reasoning behind selecting the $\lambda$ hyperparameters for G-SAN training.
The central idea here is to prioritize the loss terms based on their significance in achieving stain-morphology disentanglement.
To this end, we assign the highest value to $\lambda_{cc}$ since minimizing $\calL_{cc}$ is critical for ensuring that the stain profile and the morphology can be disentangled and put back together through the cyclic reconstruction process without any loss of information.
Similarly, to avoid the trivial solution where all the useful information is solely encoded in the morphology representation, we assign a large value to  $\calL_{lr}$ as well.
Giving the network the ability to recover the random stain vector $\vz_s^r$ that was used to produce the augmented output $\mI^{out}$ ensures that $\vz_s^r$ meaningfully contributes to the synthesized image.
\new{The effects of ablating $\calL_{lr}$ are visually presented in \cref{fig:lr_ablation}.}

The remaining loss terms in G-SAN training serve primarily to regulate the process and are thus assigned less weight. 
For instance, $\calL_{sp}$ ensures that the structural information is preserved halfway through the cyclic reconstruction process. 
However, overly emphasizing this term can limit the stain diversity in the augmented images. 
Similarly, $\calL_{id}$ and $\calL_{vae}$ are vital to SMN's formulation as a VAE. 
Still, they are not as crucial in achieving stain-morphology disentanglement and are therefore given less weight than $\calL_{cc}$ and $\calL_{lr}$.

\new{Finally, using the same nucleus segmentation experimental setup, \cref{tbl:lambda} quantitatively illustrates the effects of the various loss terms discussed above.
All losses meaningfully contribute to the performance of G-SAN.}


\begin{table}[t]
\centering
\caption{
\new{Ablation studies on several loss terms of G-SAN using the nucleus segmentation experiment.
The specific $\lambda$ values used in training our default model for this ablation study, $\text{G-SAN}_{K=3}$ @ $k=0$ or G-SAN for short, are given in the first paragraph of \cref{sec:exp}.}
}
\label{tbl:lambda}
\begin{tabular}{@{}lccc@{}}
\toprule
\multicolumn{1}{c}{Average Score} & PQ & AP & AJI \\ \midrule
$\text{G-SAN}_{K=3}$ @ $k=0$ & 0.4885 & \textbf{0.5034} & \textbf{0.4693} \\
G-SAN w/ $\lambda_{cc} = 0$ & 0.4780 & 0.4852 & 0.4548 \\
G-SAN w/ $\lambda_{lr} = 0$ & 0.4758 & 0.4885 & 0.4590 \\
G-SAN w/ $\lambda_{sp} = 0$ & 0.4711 & 0.4740 & 0.4480 \\
G-SAN w/ $\lambda_{ms} = 0$ & 0.4742 & 0.4815 & 0.4492 \\ 
G-SAN w/ $\lambda_{id} = 0$ & 0.4861 & 0.5002 & 0.4640 \\
G-SAN w/ $\lambda_{vae} = 0$ & \textbf{0.4895} & 0.4996 & 0.4659 \\ \bottomrule
\end{tabular}
\end{table}

\subsection{Novelty Comparing to Fan \etal}
\label{subsec:novelty}
\new{In this section, we discuss the fundamental differences between our G-SAN and the work by Fan \etal \cite{fan2022fast}, which also utilizes LP representation for fast stain transfer.
Most importantly, their architecture, which is almost identical to \cite{liang2021high}, is not designed for stain-morphology disentanglement and therefore is not capable of transferring to an arbitrary stain.
Furthermore, to highlight some specific yet significant differences in design, first we choose not to employ the progressive upsampling pathways, which were observed to generate undesired artifacts in the LP images in our experiments.
And second, we deliberately avoid the utilization of the ``skip-connections'' from the input BP image to the pixel-wise multiplication operator that are used in \cite{fan2022fast}.
The reason for this choice is to ensure the removal of any stain-related information from the input BP image before applying a new style, as the presence of such connections would lead to the leakage of the original image's stain into the generated image, hindering adequate stain-morphology disentanglement.}

\section{Conclusions}
In this paper, we introduced G-SAN as a domain-independent approach to stain augmentation for H\&E-stained histological images. 
By disentangling the morphological and the stain-related representations, G-SAN is capable of augmenting an input cell image with random yet realistic stains.
Additionally, by targeting the structure-preserving nature of stain transfer with a Laplacian Pyramid based architecture, the proposed G-SAN generator is highly competitive in terms of computational efficiency. 
Through the downstream tasks of patch classification and nucleus segmentation, we demonstrated quantitatively that the quality of G-SAN-augmented images is superior to the images produced by the existing stain augmentation approaches.

\bibliographystyle{IEEEtran}
\bibliography{egbib}

\end{document}